\documentclass[11pt]{article}
\usepackage{graphicx}
\usepackage{amsmath}
\usepackage{amssymb}
\usepackage{mathrsfs} 
\usepackage{amsthm}
\usepackage{tikz}
\usetikzlibrary{arrows}
\usepackage{pgfplots}
\usetikzlibrary{intersections, pgfplots.fillbetween, shapes, arrows,decorations.pathmorphing, patterns,cd}
\tikzstyle{snakeline} = [decorate, decoration={pre length=0.1cm,
                         post length=0.1cm, snake, amplitude=.5mm,
                         segment length=2mm},thick, red, ->]
\tikzstyle{arrow} = [->]

\usepackage[most]{tcolorbox}

\tcbset{colback=white!10!white, colframe=red!50!black, 
        highlight math style= {enhanced, 
            colframe=red,colback=red!10!white,boxsep=0pt}
        }

\usepackage{slashed}

\newcommand{\pa}{\partial}

\newcommand{\Tr}{{\rm Tr}}

\renewcommand{\a}{\alpha}

\newcommand{\G}{\Gamma}

\newcommand{\s}{\sigma}

\newcommand{\la}{\lambda}

\newcommand{\ud}{\mathrm{d}}          
\newcommand{\ue}{\mathrm{e}}    
\newcommand{\be}{\begin{equation}}
\newcommand{\ee}{\end{equation}}
\newcommand{\bea}{\begin{eqnarray}}
\newcommand{\eea}{\end{eqnarray}}
\newcommand{\ba}{\begin{align}}
\newcommand{\ea}{\end{align}}

\newcommand{\nn}{\nonumber\\}

\newcommand{\cL}{\mathcal{L}}

\newcommand{\cM}{\mathcal{M}}
\newcommand{\cH}{\mathcal{H}}
\newcommand{\cO}{\mathcal{O}}
\newcommand{\cF}{\mathcal{F}}

\newcommand{\cC}{\mathcal{C}}
\newcommand{\cD}{\mathcal{D}}

\newcommand{\cK}{\mathcal{K}}

\newcommand{\cR}{\mathcal{R}}

\usepackage{jheppub}

\title{\boldmath Conformal quantum mechanics as a Floquet-Dirac system}

\author{Rodrigo de Le\'on Ard\'on }

\affiliation{ Universidad Andr\'es Bello,\\ Departamento de Ciencias F\'isicas, Facultad de Ciencias Exactas, 
 Sazi\'e 2212, Piso 7, Santiago, Chile.}
\emailAdd{rodrigoardon.e@gmail.com}

\abstract{Conformal quantum mechanics has been proposed to be the CFT$_1$ dual to AdS$_2$. 
The $N$-point correlation function  that satisfy conformal constraints have been constructed from a non-conformal vacuum and the insertion of a non-primary operator. The main goal of this paper is to find an interpretation of this oddness. For this purpouse, we study possible gravitational dual models and propose  
a two-dimensional dilaton gravity with a massless fermion for the description of conformal quantum mechanics. We find a universal   correspondence between states in the conformal quantum mechanics model and two-dimensional spacetimes. Moreover, the solutions of the Dirac equation can be interpreted as zero modes of a Floquet-Dirac system. Within this system, the oddness of the non-conformal vacuum and non-primary operator is  elucidated. As a possible application, we interpret the  gauge symmetries of the Floquet-Dirac system as the corresponding  infinite symmetries of the Schr\"odinger equation which are conjectured to be related to higher spin symmetries.
}
\begin{document} 
\maketitle
\flushbottom

%
%
%
%
%
%
%
%
%

\section{Introduction}
 Conformal quantum mechanics, CQM, is a subject continuously studied\footnote{A list of papers are detailed in the references of \cite{Gonera:2012gr,Andrzejewski:2015jya}.} since its birth \cite{Jackiw:1972cb, D'Alfaro:693633}. Physical applications range from systems such as magnetic monopoles, magnetic vortex, molecular dynamics, neutral atoms and charged wires, Efimov effect, black holes, membranes to quantum cosmology, see e.g. \cite{JACKIW1980183,JACKIW199083,Camblong:2001zt,Coon:2002sua,Camblong:2003mz,Naidon:2016dpf,Claus:1998ts,BrittoPacumio:1999ax,Okazaki:2015pfa,Pioline:2002qz,BenAchour:2019ufa} and references within. In the light of the AdS/CFT duality \cite{Maldacena:1997re,Gubser:1998bc,Witten:1998qj} , CQM has been proposed to be the CFT$_1$ dual to AdS$_2$ in \cite{CHAMON2011503,Jackiw:2012ur}. The key message explained in \cite{CHAMON2011503}, based on previous results given in \cite{D'Alfaro:693633}, is that correlation functions of primary operators that satisfy the conformal structure can be built from a non-conformal vacuum, the so called $R$-vacuum, and an operator that is not primary. 

The reason of this oddness is attributed to the fact that in CQM we are dealing with a Hilbert space rather than a Fock space as in any $d\geq2$ CFT. In $d=1$ one cannot construct a normalized vacuum state that is annihilated by all the generators of the conformal group. It turns out that the combination of  the $R$-vacuum with the non-primary operator conspire in the correlation functions to respect conformal constraints. This can be taken to be just a particularity in the AdS$_2$/CQM dictionary. However, since the system corresponds to the simplest playground to study AdS/CFT, further analysis must be done in order to give more sense of the aforementioned oddness.

On the other hand, since CQM can be realized in physical systems, such analysis can be always contrasted with physical reasoning. This may take an orthogonal direction of \cite{CHAMON2011503,Jackiw:2012ur}, since they are based on symmetries rather than a specific model. Nevertheless, one of the goals of this work is to complement their conclusions. 

Concretely, taking the CQM realization to be the dipole-charge system as studied in \cite{Camblong:2001zt}, one deals with a radial inverse square potential. The three-dimensional Schr\"odinger equation reduces to one-dimensional CQM model defined in \cite{D'Alfaro:693633}. In \cite{PhysRevD.48.5940,Camblong:2000ec,Camblong:2003mb,PhysRevD.68.125013,doi:10.1119/1.2165248, Gitman_2010} and references within, it has been shown that the attractive radial inverse square potential problem leads to self-adjoint extensions, the notion of renormalization and anomalies in CQM. This is intuitive since  a bound state arising from the attractive potential  implies a physical scale and therefore this will break scale invariance at the quantum level. Regarding the self-adjoint extensions, ambiguities arise for the adequate boundary conditions.  Interestedly, these and other subjects has been addressed in \cite{Burgess:2016lal} from an effective field theory perspective.

However, these results should  be contrasted with \cite{Andrzejewski:2015jya}, where a fully conformally invariant quantum mechanics for all values of the coupling constant has been constructed. The main argument is that the notion of quantum anomalies can be traced to the non-trivial topology of classical phase-space. Therefore, once proper covering of the phase space is constructed the dynamics becomes smooth and the quantization can be constructed.

In \cite{D'Alfaro:693633} some of the above issues do not arise since their analysis is based on the repulsive inverse square potential problem. Nevertheless, the one-dimensional CQM model faces the intertwined issue that the Hamiltonian $H$ is not self-adjoint and  the ground state is not normalizable. This is overcome by means of the conformal structure. A compact operator $R$ is chosen for the evolution of the system. The new lowest eigenstate is normalizable and is referred as to the $R$-vacuum. The same path is taken in \cite{CHAMON2011503,Jackiw:2012ur}, where the $R$-vacuum corresponds to the lowest state of the representation.

One can take another approach regarding the issue of having non-normalizable vacuum state as discussed in \cite{Okazaki:2017lpn}. Not having a normalizable vacuum state does not imply the absence of a physically well-defined description. It rather indicates that the quantization needs to take into consideration possible constraints on the canonical variables, see e.g. \cite{Strocchi:2016kce}. In the case of the inverse square potential, the constraint is realized classically as the restriction to half phase-space\footnote{This is close related to \cite{Andrzejewski:2015jya}.}. In fact an interpretation of this ``ill-defined'' Hamiltonian is found in \cite{10.1093/ptep/pty058}.

Hence, the inverse square potential as a realization of CQM is by far not a trivial model. From an AdS/CFT perspective, a natural question arises: what is the holographic description of this potential that encompass all of its properties?
The results of \cite{D'Alfaro:693633,CHAMON2011503,Jackiw:2012ur} indicates a dual description for the repulsive case. This avoids the discussion regarding anomalies associated with bound states. As mentioned,  the results in \cite{CHAMON2011503,Jackiw:2012ur} are independent of a particular model to CQM\footnote{Works in these lines are vast. Consider e.g. \cite{PhysRevD.92.126010,Pinzul:2017wch,deAlmeida:2019awj} and references within.} and the general prescription for the $N$-point function is given. These functions are in agreement with the conformal structure and are constructed in terms of a non-conformal vacuum and the insertion of non-primary operator. Moreover, these $N$-point functions should correspond to the ones computed in AdS$_2$.

If this is the case, two main questions arises:
\begin{itemize}
\item The computation in the AdS$_2$ side, using the usual prescription, involves a conformal invariant vacuum. The CFT$_1$, in the view of \cite{CHAMON2011503,Jackiw:2012ur},  does not. Is the $R$-vacuum and a non-primary operator, together with their conspiracy to respect the conformal structure, realized in a gravitational setup?

\item What is the physical interpretation of the non-primary operator?
\end{itemize}

This work is devoted to answering these questions. It is organized as follows: in section II we review the CQM model developed in \cite{D'Alfaro:693633} and its application to AdS/CFT  developed in \cite{CHAMON2011503,Jackiw:2012ur}. In section III, partially motivated by \cite{Claus:1998ts,BrittoPacumio:1999ax,Pioline:2002qz} we undertake a geometrical exploration of possible dual models. The exploration is based on the $R$ operator expressed in two bases: $\{|q\rangle\}$ and $\{|t\rangle\}$. For the former, the eigenvalue problem of the operator is obtained by considering the canonical quantization of a worldline in a negative curved spacetime. Moreover, we can interpolate between the Hamiltonian of the system and the  $R$ operator by tuning the corresponding Wheeler-DeWitt equation of the worldline. The Hamiltonian case turns out to be the massless/tensionless limit.  If the static background is considered to arise from the near-horizon geometry of a Reissner-Nordstr\"om black hole, the mass and charge must satisfy $\frac{2}{3}\sqrt{2}M<Q< M$. Another possibility with less fine tuning is that the background corresponds to a solution of a two-dimensional dilaton gravity in the weak gravity limit.

For the $|t\rangle$ basis, the geometrical description does not requires a worldline but rather a fermion. The equation of the $R$ operator is obtained from the Dirac equation in a non-static background. This yields a complex geometry and a general interpretation of the $N$-point function in terms of the fermion. Moreover, the solutions of the Dirac equation can be interpreted as zero modes of a Floquet-Dirac system. Within this interpretation, the main questions are answered in great detail.

In section IV we propose a general dual description and conclusions. The general dual model is based on the results of the exploration and consist of a two-dimensional dilaton gravity with a massless fermion. The equation of motion of two distinct dimensionally reduced systems  correspond to the action of the $R$ operator in the $\{|q\rangle\}$ and $\{|t\rangle\}$ basis. Finally, we give a summary of the relation between the parameters of the CQM and  Floquet-Dirac models. As an application, we discussed the possible interpretation of the gauge  symmetries of the Floquet-Dirac system  as the corresponding infinite symmetries of the Schr\"odinger equation which are conjectured to be related to higher spin symmetries.

\section{CQM review}
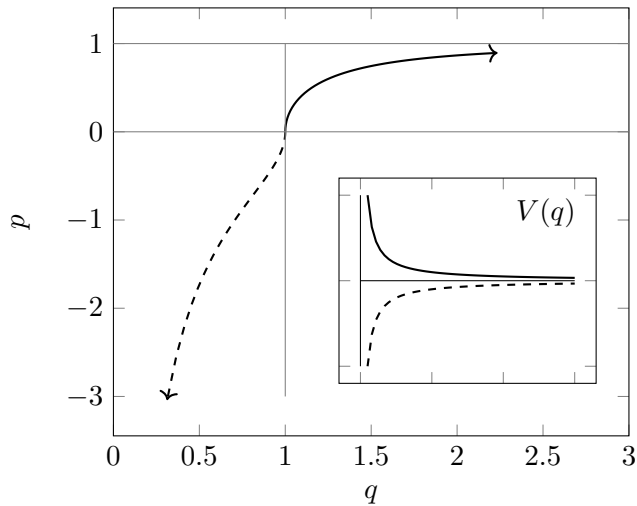
\begin{figure}[ht!]
\centering
  \begin{tikzpicture}
\begin{axis}[
	xlabel={$q$},
    ylabel={$p$},
    xmin=0,xmax=3,
]
\addplot [->,domain=1:3, samples=50,thick] ({sqrt((x-1)^2+1)},{(x-1)/sqrt((x-1)^2+1) });
\addplot [->,domain=1:1.95, samples=50, dashed, thick] ({sqrt(-(x-1)^2+1)},{-(x-1)/sqrt(-(x-1)^2+1) });
\addplot[color=gray] coordinates {(0,0) (3,0)};
\addplot[color=gray] coordinates {(0,1) (3,1)};
\addplot[color=gray] coordinates {(1,-3) (1,1)};
\end{axis}
  \begin{axis}[
	xshift=3cm,
	yshift=0.7 cm,
	width=5cm,	
	yticklabels={,,},
   xticklabels={,,},
]
\addplot [domain=0.1:3, samples=50,thick] ({x},{1/x });
\addplot [domain=0.1:3, samples=50,dashed,thick] ({x},{-1/x });
\addplot[color=black] coordinates {(0,0) (3,0)};
\addplot[color=black] coordinates {(0,-10) (0,10)};
\node at (260,180) {$V(q)$};
 \end{axis}

\end{tikzpicture}
\caption{Classical solutions in phase space for $t\geq t_0$ and $q(t_0)=1$. The dashed curve corresponds to $g<0$ (attractive case) and the regular curve to $g>0$ (repulsive case). In the subfigure, the potential is shown for the repulsive and attractive cases. }
\label{conformalparticle}
\end{figure}

The departure point is  the conformal quantum mechanics model given in  \cite{D'Alfaro:693633}. The classical Hamiltonian of the model is given by 
\be 
H(p,q)=\frac{1}{2}p^2+\frac{g}{2}\frac{1}{q^2},
\ee
 where $g$ is a dimensionless coupling constant.  The equations for the canonical variables are $\dot{p}=\frac{g}{q^3}$ and $\dot{q}=p$. Combining the equations we obtain $\ddot{q}=\frac{g}{q^3}$. The solutions are 
\bea 
q(t)&=&\sqrt{v^2(t-t_0)^2+q^2(t_0)}, \label{qeq}\\
p(t)&=&\frac{v^2(t-t_0)}{\sqrt{v^2(t-t_0)^2+q^2(t_0)}},
\eea
where $v^2=\frac{g}{q^2(t_0)}$. For $g>0$ ($v^2>0$) and $v^2(t-t_0)^2\gg q^2(t_0)$ we obtain $q\sim vt$ and $p\sim v$. On the other hand, for $g<0$ ($v^2<0$) and $|v^2|(t-t_0)^2\to q^2(t_0)$ we obtain $p\to-\infty$ and $q\to 0$ . These regimes are illustrated in Figure \ref{conformalparticle}. 

For the repulsive case, $g>0$, the particle is asymptotically free. For the attractive case, the particle reaches the $q=0$ singularity (the boundary in phase-space) in a finite time. This can be calculated from Eq. \eqref{qeq} to be $\Delta t=\frac{q^2(t_0)}{\sqrt{|g|}}$. 

At the quantum level, the Schr\"odinger equation is
\be
 \frac{1}{2}\left[-\frac{\ud^2}{\ud q^2}+\frac{g}{q^2}\right]\psi=E\psi.
 \label{SeqH}
\ee
As discussed in \cite{doi:10.1119/1.2165248,Gitman_2010}, $H$ can be Hermitian (or symmetric) but not self-adjoint. Let $\phi$ and $\psi$ be two square integrable functions that vanish, together with their first derivative, at $q\to\infty$. Then,
\be
 \langle \phi |H\psi\rangle =\langle H\phi|\psi\rangle+\frac{1}{2}\left.\left(\phi^*\frac{\ud\psi}{\ud q}-\psi\frac{\ud\phi^*}{\ud q}\right)\right|_{q=0}.
\ee
We further restrict $\psi$ to vanish, together with its first derivative, at $q=0$. Then, we find that $H$ is Hermitian in the respective domain. Notice that the domain of the adjoint is larger since $\phi$ and its derivative are not restricted at $q=0$. Therefore, $H$ is not self-adjoint.

On the other hand, the ground state wave function is
\be
\psi_0(q)=\left\{\begin{matrix}
c_+q^{\frac{1}{2}(1+\sqrt{1+4g})}+c_-q^{\frac{1}{2}(1-\sqrt{1+4g})}&g\neq -\frac{1}{4}\\
c_1q^{\frac{1}{2}}+c_2q^{\frac{1}{2}}\ln (q/a)& g=-\frac{1}{4}
\end{matrix}\right. .
\label{groundH}
\ee
The logarithmic solution breaks scale invariance due to the presence of the scale $a$. The condition that the wave function and its derivative must vanish at $q=0$ can only be applied for the $g\neq -\frac{1}{4}$ case. It corresponds to demanding  $\frac{1}{2}\pm\frac{1}{2}\sqrt{1+4g}>0$ and $-\frac{1}{2}\pm\frac{1}{2}\sqrt{1+4g}>0$. Only the positive root can satisfy both restrictions for $g>0$. The problem with the resulting solution, together with its first derivative, is that it diverges as $q\to \infty $. These intertwined issues are encountered in \cite{D'Alfaro:693633} for the repulsive case $g>0$. The study for other values of the coupling and the self-adjoint extensions has been carried out in great detail in \cite{doi:10.1119/1.2165248,Gitman_2010} and discussed in \cite{Camblong:2001zt,Andrzejewski:2015jya}.

In order to deal with the issues for the repulsive case one should notice, as in \cite{Jackiw:1972cb, D'Alfaro:693633,JACKIW1980183,JACKIW199083}, that the classical equations of motion are invariant under 
\bea
t'&=&\frac{at+b}{ct+d},\label{timerep}\\
q'(t')&=&\frac{1}{ct+d}q(t),\\
p'(t')&=& (ct+d) p(t)-cq(t),
\eea
where $a,b,c,d\in\mathbb{R}$ and $ad-bc=1$. After arranging the parameters into a matrix form 
\be
G=\begin{pmatrix}
a&c\\
b&d
\end{pmatrix},
\ee
and considering two successive transformations over $t$ we see that $G$ is an element of the group $SL(2,\mathbb{R})$. It is well known that there is a two-to-one homomorphism between $SL(2,\mathbb{R})$ and the conformal group in one dimension $SO(2,1)$. Therefore,  we found that our system (after restricting some parameters) is conformally invariant. We can see this by noticing that the transformation includes time evolution (parametrized by $b$ with $a=d=1$, $c=0$), dilatations  (parametrized by $a$ with $d=1/a$, $c=b=0$) and special conformal transformations (parametrized by $c$ with $a=d=1$, $b=0$). 

These three transformations can be interpreted as one-parameter family of canonical transformations. As known, the generator of time evolution is $H$. For dilatations we set $a=1+\frac{\a}{2}$ where $\a$ is small. Then,
\be
q'(t)-q(t)=-\a\{q,Ht-\frac{1}{2}pq\},\quad
 p'(t)-p(t)=-\a\{p,Ht-\frac{1}{2}pq\}.
\ee
Therefore the generator of dilatations corresponds to
\be 
D(p,q,t)=Ht-\frac{1}{4}(pq+qp).
\ee
Similarly, the generator of special conformal transformations  is defined as
\be 
K(p,q,t)=Ht^2-\frac{1}{2}(pq+qp)t+\frac{1}{2}q^2. 
\ee
One can verify that $D(p,q,t)$ and $K(p,q,t)$ are actually constants of motion. Moreover,  $\cH$, $\cD$ and $\cK$ satisfy 
\be
\{ D, H\}=- H,\,\,\{ D, K\}= K,\,\,\{ K, H\}=-2 D,    
\label{PoissonAlgebra}
\ee
where $\{,\}$ is the Poisson bracket. This is the explicit realization of the $\mathfrak{sl}(2,\mathbb{R})$ algebra.

The time independent generators  $H$, $D_0=D(p,q,0)$ and $K_0=K(p,q,0)$ naturally also satisfy  the $\mathfrak{sl}(2,\mathbb{R})$ algebra. After its quantization, i.e. $\{\cdot,\cdot\}\to\frac{1}{i}[\cdot,\cdot]$, 
the algebra in the Cartan-Weyl basis reads
\be
[R,L_{\pm}]=\pm L_{\pm}, \quad [L_-,L_+]=2R, 
\ee
where 
\be 
R=\frac{1}{2}\left(\frac{1}{t_*}K_0+t_*H\right),\,\,L_{\pm}=\frac{1}{2}\left(\frac{1}{t_*}K_0-t_*H\right)\pm iD_0,
\ee
 where $t_*$ is introduced by dimensional reasons and it is set to unity for convenience. 
 
As discussed in \cite{D'Alfaro:693633,CHAMON2011503}, the generator $R$ is compact and therefore its spectrum is discrete. In order to understand this statement, we adopt the Schr\"odinger picture. The  eigenvalue problem of the generator $R$ result
\be
\frac{1}{4}\left[-\frac{\ud^2}{\ud q^2}+\frac{g}{q^2}+q^2\right]\varphi=r \varphi.
\label{SeqR}
\ee
The ``effective'' potential is $V\sim \frac{g}{q^2}+q^2$ and due to the presence of the $q^2$ term, $R$ is bounded from below.  Notice the ``effective'' potential corresponds to a three-dimensional isotropic harmonic oscillator\footnote{We refer to the reader to \cite{Inzunza:2019sct,Inzunza:2021fep,Inzunza:2021vgt} for a detail discussion of transformations (including their proposal called ``conformal bridge transformations'') between quantum systems that enjoy conformal symmetry.  }.

Equation \eqref{SeqR} can be factorized by means of the superpotential, see table 4.1 in \cite{doi:10.1142/4687},
\be
W(q)=q-\frac{2r_0-\frac{1}{2}}{q}, \quad r_0=\frac{1}{2}\left(1\pm\sqrt{\frac{1}{4}+g}\right),
\ee
where $r_0\geq \frac{3}{4}$\footnote{This expression is obtained from the comparison between $l$ (the azimuthal quantum number of the three-dimensional isotropic harmonic oscillator) and $r_0$. They are related by $2r_0=\frac{3}{2}+l$. Since $l\geq 0$ we obtain $r_0\geq \frac{3}{4}$. }.  This restriction on $r_0$ can only be satisfied by the positive root of $r_0$ and $g>0$. From the creation and annihilation operators
\be
A^{\dagger}=-\frac{\ud}{\ud q}+W(q),\quad A=\frac{\ud}{\ud q}+W(q),
\ee
we obtain $(4R-4r_0)\varphi_n=(4r_n-4r_0)\varphi_n=A^{\dagger}A\varphi_n=4n\varphi_n$. Thus $r_n=r_0+n$ for $n=0,1,\ldots$. The ground state is of the form 
\be
\varphi_0=C q^{\frac{1}{2}+\frac{1}{2}\sqrt{1+4g}}\ue^{-\frac{q^2}{2}}.
\label{groundR} 
\ee
Notice that the exponential factor makes the difference with respect to Eq. \eqref{groundH}. The ground state satisfies $\varphi_0(0)=0$ and $\varphi_0'(0)=0$. Therefore the only remaining issue  to address is the self-adjointess of $R$. As shown in appendix \ref{appA}, $R$ is in fact self-adjoint.

From a group theory point of view,  \cite{CHAMON2011503} arrives to the same result from 
\be
R|n\rangle =r_n|n\rangle,\quad n=0,1,\ldots,\quad r_0>0, \quad \langle n' | n \rangle =\delta_{n' n}.
\ee 
$r_0$ corresponds to the lowest weight of the representation and $|0\rangle$ is the lowest state referred as to the $R$-vacuum. The action of the ladder operators and Casimir are
\bea
L_{\pm}|n\rangle&=&\sqrt{r_n(r_n\pm 1)-r_0(r_0-1)}|n\pm1\rangle, \\
\cC|n\rangle &=&r_0(r_0-1)|n\rangle\label{casimireigenp},
\eea
where the Casimir operator is given by
\be
\cC=\frac{1}{2}(HK_0+K_0H)-D_0^2=R^2-L_+L_-. \label{casimirdef}
\ee 
The key point to stress is that for a generic $r_0$, which has to be determined from the dynamics, not all the generators of the algebra in the Cartan-Weyl basis annihilate the $R$-vacuum\footnote{ Indeed: $R|0\rangle=r_0|0\rangle$, $L_-|0\rangle=0$ and $L_+|0\rangle=\sqrt{2r_0}|1\rangle$.}. We know that  in \cite{D'Alfaro:693633} $r_0\geq \frac{3}{4}$.

 In terms of the $\{H,D_0,K_0\}$ basis, this translates to the statement that there is no state $|\Omega\rangle$ that is simultaneously annihilated by the all generators. Nevertheless, in \cite{D'Alfaro:693633,CHAMON2011503}, it was shown that in spite of the absence of a conformal invariant vacuum, correlation functions obeying conformal constraints can be constructed. 
 
The first step is to consider the representation
\be
H=i\frac{\ud }{\ud t},\quad
D_0=i\left(t\frac{\ud }{\ud t}+r_0\right),\quad
K_0=i\left(t^2\frac{\ud }{\ud t}+2r_0t\right),
\label{t-rep-sl2al}
\ee 
where $t\in\mathbb{R}$. They satisfy the (quantized) $\mathfrak{sl}(2,\mathbb{R})$ algebra together with  equations \eqref{casimirdef} and \eqref{casimireigenp}. The $R$ operator becomes
\be
R=\frac{i}{2}\left(\left(t_*+\frac{t^2}{t_*}\right)\frac{\ud }{\ud t}+2r_0\frac{t}{t_*}\right),
\label{Rt}
\ee
where we have restored $t_*$. It is useful to define the states $|t\rangle$ such that $\langle t|R|n\rangle=R\langle t|n\rangle=r_n\langle t|n\rangle$. Then, 
\be
 \langle t|n\rangle =C_n\left(\frac{t_*-it}{t_*+it}\right)^{r_0+n}\frac{1}{(1+\frac{t^2}{t_*^2})^{r_0}}.
 \label{tn1}
\ee
It is interesting to notice that 
\be
  \langle t_1|t_2\rangle= \sum_{n=0}^{\infty}\langle t_1|n\rangle\langle n|t_2\rangle,
  =\frac{t_*^{4r_0}}{[(t_*+it_1)(t_*-it_2)]^{2r_0}}\sum_{n=0}^{\infty}|C_n|^2(r(t_1,t_2))^n,\label{sum1}
\ee
where 
\be
 r(t_1,t_2) = \frac{(t_*-it_1)(t_*+it_2)}{(t_*+it_1)(t_*-it_2)}.
\ee
Using the ratio test, the sum converges if $L|r|<1$, where $L=\lim_{n\to\infty}\frac{|C_{n+1}|^2}{|C_n|^2}$. Following \cite{D'Alfaro:693633}, the constants are 
\be 
C_n=(-1)^n\sqrt{\frac{\G(n+2r_0)}{\G(n+1)}}.
\label{Cncoeff}
\ee
Then $L=1$ and for $|r|<1$ we obtain
\be
 \langle t_1|t_2\rangle=\frac{t_*^{2r_0}\G(2r_0)}{(2i(t_1-t_2))^{2r_0}}.
 \label{2pfunc}
\ee
It is intriguing that the inner product gives the two-point function expect from the CFT$_1$. This suggest that  the states $|t\rangle$ may be coherent states. But  from
\bea
\int\limits_{-\infty}^{\infty}\ud t\,| t\rangle \langle t|
&=&\sum_{n,m=0}\int\limits_{-\infty}^{\infty}\ud t\,\langle m|t\rangle\langle t|n\rangle |m\rangle\langle n|= t_*\sum_{n,m=0}C_{mn}(r_0)|m\rangle\langle n|,\nn
C_{mn}(r_0)&=&2(-1)^{1+m+n+2r_0}C_mC_n (m-n-2r_0)\Gamma(4r_0-1)\sin(2r_0\pi)\label{Cnm},
\eea
we see that  the states $|t\rangle$  do not resolve the unity and therefore the coherent interpretation must be discarded\footnote{ See e.g. \cite{gazeau2009coherent} for details on coherent states.}. However, as shown in  \cite{CHAMON2011503}, there exists a sort of ``displacement'' operator $\cO(t)$ defined as
\bea
|t\rangle &=& \cO(t)|0\rangle, \\
\cO(t) &=&N(t)\ue^{-\omega(t)L_+},\\
N(t)&=&\sqrt{\G(2r_0)}	\left(\frac{\omega(t)+1}{2}\right)^{2r_0},\\
\omega(t)&=&\frac{t_*+it}{t_*-it}.
\eea
Therefore, the two-point function corresponds to $\langle 0|\cO^{\dagger}(t_1)\cO(t_2)|0\rangle$. As stated in \cite{CHAMON2011503},  the $R$-vacuum is not conformally invariant and $\cO(t)$ is not a primary but the combination conspire to give a conformal two-point function. The general prescription for the $N$-point function is given \cite{CHAMON2011503,Jackiw:2012ur}. Let $O_{\Delta}(t)$ be a primary operator, then the $N$-point function is of the form
\be
\langle t_1|O_{\Delta}(t_2)\cdots O_{\Delta}(t_{N-1})|t_N\rangle= \langle 0|\cO^{\dagger}(t_1) O_{\Delta}(t_2)\cdots O_{\Delta}(t_{N-1})\cO(t_N)|0\rangle.
\ee
Moreover, these $N$-point functions should correspond to the ones computed in $AdS_2$.

\section{Geometric description: an exploration}
\subsection{Worldine}
As a warm up, let us consider a worldline in $AdS_2$ in the Poincar\'e patch. The metric is given by
\be
\ud s^2=\frac{L^2}{Q^2}(-\ud T^2+\ud Q^2),
\ee
where $L$ is the radius of $AdS_2$. For $d=2$ we have two boundaries, we restrict the analysis for one of them. The action is considered to be
\be
S=-M\int\ud\la\sqrt{-g_{\mu\nu}(X)\dot{X}^{\mu}\dot{X}^{\nu}},
\label{worldlineaction}
\ee
 where $M>0$ is a mass parameter, $X^{\mu}=(T,Q)$ and $\dot{} =\frac{\ud}{\ud \la}$. Due to reparametrization invariance, the Hamiltonian of the system is constrainted to vanish \cite{Henneaux:1992ig}
\be
H\equiv g^{\mu\nu}(X)P_{\mu}P_{\nu}+M^2=0,\quad P_{\mu}=\frac{-M g_{\mu\nu}\dot{X}^{\nu}}{\sqrt{-g_{\mu'\nu'}(X)\dot{X}^{\mu'}\dot{X}^{\nu'}}}. 
\label{Hconst}
\ee
In the Schr\"odinger picture, after choosing the ordering of the operators displayed in Eq. \eqref{Hconst}, the constraint becomes the Wheeler-DeWitt equation 
\be
\left(\frac{\pa^2}{\pa T^2}-\frac{\pa^2}{\pa Q^2}+\frac{M^2L^2}{Q^2}\right)\Psi=0.
\label{WdWeq1}
\ee
By writing the wave function as $\Psi=\ue^{\pm i\omega T}\psi(Q)$ we obtain $-\psi''+(ML)^2/Q^2 \psi=\omega^2\psi$. 

After setting $g=\frac{M^2L^2}{2}$ and $E=\frac{\omega^2}{2}$, we recover Eq. \eqref{SeqH}. The relation between the Wheeler-DeWitt equation and the Sch\"odinger equation is also found in \cite{Pioline:2002qz}. For our case, it  should not be a surprise since the group $SL(2,\mathbb{R})$ is a subgroup of the isometries of the spacetime. Moreover, the action is reparametrization invariant and the conformal transformation given in Eq. \eqref{timerep} corresponds to a subcase.

The interesting point is that the potential emerges from the conformal factor. Notice that $g>0$ and therefore the worldline describes the repulsive case. Moreover, we see that the Euclidean worldline\footnote{We set $T=-iT_E$, the resulting equation is obtained by the substitution of  $M^2\to-M^2$ in Eq. \eqref{WdWeq1}.} will cover the attractive case. The $T$-dependence on the wave function shows the difference between the two cases: it oscillates for $g>0$ and exponentially decays/grows for $g<0$.

Another model can be constructed based on the results given in \cite{Pioline:2002qz}. Consider the Polyakov action
\be
S=\int\ud\la\, \cL,\quad \cL=\frac{1}{e}G_{IJ}(Y)\dot{Y}^I\dot{Y}^J+e\Lambda , 
\label{WLPol}
\ee
where $e$ is an auxiliary field and $G_{IJ}(Y)$ corresponds to the metric of a $d$-dimensional spacetime $\cM$. Substituting the equation of motion of $e$ we obtain
\be
S=2\sqrt{\Lambda}\int\ud\la\sqrt{G_{IJ}(Y)\dot{Y}^I\dot{Y}^J}. 
\ee
The momenta computed from Eq. \eqref{WLPol} are
\be
P_I=\frac{\pa \cL}{\pa \dot{Y}^I}=\frac{2}{e}G_{IJ}(Y) \dot{Y}^J,\quad \Pi =\frac{\pa \cL}{\pa \dot{e}}=0.
\ee
We see that $\Pi=0$ corresponds to a primary constraint. The Hamiltonian result 
\be
\cH=\frac{e}{4}H,\quad H=G^{IJ}(Y)P_IP_ J-4\Lambda. 
\ee
The consistency condition $\dot{\Pi}=0$ implies that $H$ is a secondary constraint and applying again the consistency condition no further constraint is found. The system is first class. Let us assume now that the metric is of the form
\be
G_{IJ}(Y)\ud Y^I\ud Y^J=f(q)\ud q^2 +h(q)g_{\mu\nu}(y)\ud y^{\mu}\ud y^{\nu}, \quad q\geq 0.
\ee
Then, 
\be
H=\frac{1}{f(q)}\left[P_q^2+\frac{f(q)}{h(q)}g^{\mu\nu}(y)P_{\mu}P_{\nu}-4\Lambda f(q)\right]. 
\label{Hfh}
\ee
At the quantum level, the wave function $\Psi$ must satisfy the quantum constraints $\hat{\Pi}\Psi=0$ and $\hat{H}\Psi=0$. In the Sch\"odinger picture we obtain
\be
\frac{\pa \Psi}{\pa e}=0,\quad  \left[-\frac{\pa^2 }{\pa q^2}-\frac{f(q)}{h(q)}\Box_y-4\Lambda f(q)\right]\Psi=0,
\label{quantumconstraints}
\ee
where for the second expression, the Wheeler-DeWitt equation, we follow the ordering as displayed in Eq. \eqref{Hfh} and $\Box_y=g^{\mu\nu}(y)\pa_{\mu}\pa_{\nu}$. The first part of Eq. \eqref{quantumconstraints} indicates that $\Psi$ is independent of $e$. For the Wheeler-DeWitt equation we write $\Psi(q,y)=\psi(q)\Theta(y)$ and assume $\Box_y\Theta=-\frac{g}{L^2}\Theta$ where here $L$ is an intrinsic length scale. The final equation result
\be
  \left[-\frac{\ud^2 }{\ud q^2}+\frac{f(q)}{h(q)}\frac{g}{L^2}-4\Lambda f(q)\right]\psi=0.
  \label{EqSCH}
\ee
Let $q=L\bar{q}$ and $\Lambda=\frac{\bar{\Lambda}}{L^2}$, where $\bar{q}$ and $\bar{\Lambda}$ are dimensionless. The line element of this spacetime $\cM$ is
\be
\ud s^2= L^2f(L\bar{q})\ud \bar{q}^2 +h(L\bar{q})g_{\mu\nu}(y)\ud y^{\mu}\ud y^{\nu}. 
\label{EqLineseg}
\ee
From the equations \eqref{EqSCH} and \eqref{EqLineseg} we see that $f$, $h$ and $g_{\mu\nu}$ must be dimensionless. By choosing
\be
 f(q)=\frac{q^2}{L^2}=\bar{q}^2,\quad h(q)=\frac{q^4}{L^2}\frac{1}{L^2-\bar\a q^2}=\frac{\bar{q}^4}{1-\bar\a \bar{q}^2},
\ee
we obtain 
\be
  \left[-\frac{\ud^2 }{\ud \bar{q}^2}+\frac{g}{\bar{q}^2}-4\bar\Lambda \bar{q}^2\right]\psi=g\bar\a\psi.
  \label{ResultingWDW}
\ee
If we further choose $\bar\Lambda=-\frac{1}{4}$ and $\bar{\a}=\frac{4r}{g}$,  the dimensionless eigenvalue problem of the $R$-operator given in Eq. \eqref{SeqR} is recovered. Let us set $d=2$ and $g_{\mu\nu}(y)\ud y^{\mu}y^{\nu}=-\ud y^2$. Therefore $\Theta_{\pm}(y)=\Theta_0\exp(\pm\frac{\sqrt{g}y}{L})$. Notice that $G_{IJ}$ has Lorentzian signature and therefore the line segment action result $S=-\int\ud\la\sqrt{-G_{IJ}(Y)\ud Y^I\ud Y^J}$. The minus sign arises from the square root terms. In the massless/tensionless limit, defined as $\Lambda\to 0$, we recover the dimensionless version of Eq. \eqref{SeqH} for $\bar{\a}=\frac{2\bar{E}}{g}$.

The resulting spacetime line segment is
\be
\ud s^2=\frac{q^2}{L^2}\left[ -\frac{q^2}{L^2-\bar\a q^2}\ud y^2 +\ud q^2\right], \quad \bar\a\geq 0.
\label{EqLineseg2}
\ee
In order to study the signature we study the behavior of the function 
\be
F(q)= \frac{q^2}{L^2-\bar\a q^2}.
\ee
The function is discontinuous at $q_*=\frac{L}{\sqrt{\bar{\a}}}$. For $0\leq q<q_*$ we have $F\geq 0$  and $F<0$ for $q>q_*$. In order to fix the Lorentzian signature we push  the ``wall'' at $q_*$ to infinity, i.e. we demand $\bar{\a}\to 0$. Since $\bar\a=\frac{4r}{g}=\frac{4r_0+4n}{g}$ and $r_0\geq\frac{3}{4}$, we obtain
\be
 \frac{\sqrt{g}}{\sqrt{3+4n}}\to \infty.
\ee
Therefore $\sqrt{g}\gg \sqrt{3+4n}$ for all $n$. For large and finite $g$, the geometrical description will be accurate below some $n_c$. In particular for the ground state we only demand that $g\gg 3$.

Notice that we have a collection of spacetimes $\cM_n$, in principle one per each state $|n\rangle$. In order to study their properties we find the Ricci scalar to be
\be
\cR_n= -\frac{2L^2\bar\a_n(4L^2-\bar{\a}_n q^2)}{q^2(L^2-\bar\a_n q^2)^2}=-\frac{2}{L^2}\cF(\bar\a_n,\bar{q}),\quad \cF(\bar\a_n,\bar{q})=\frac{\bar\a_n(4-\bar{\a}_n \bar{q}^2)}{\bar{q}^2(1-\bar\a_n \bar{q}^2)^2}.
\label{riccia}
\ee
Notice that $\cF$ diverges at $\bar{q}=\{0,\frac{1}{\sqrt{\bar\a_n}}\}$ and is positive and bounded from below only between these points. The second ``wall'' is located at the same place of the one in $F$ and in this interval $\cF(\bar{q}_*=\sqrt{3-\sqrt{7}}\frac{1}{\sqrt{\bar\a_n}})=\frac{1}{3}(37+14\sqrt{7})\bar\a_n^2$ is a global minimum. Therefore, as we push the ``wall'' to infinity the Ricci scalar asymptotically goes to zero. Hence the spacetimes have negative curvature and they are asymptotically flat. The singularity at $q=0$ is expected and therefore whether it is a true singularity or not is not relevant to our discussion.

For the massless/tensionless case we see that $\bar\a=0$ for the ground state $E=0$. We find that $F\geq 0$ and therefore the Lorentzian signature is fixed and the manifold is flat. In fact, it corresponds to a right Rindler wedge after the coordinate transformation $q=L\ue^{\frac{\xi}{L}}$. The proper acceleration is $\frac{2}{L}$. In this case, to each continuous energy there is a corresponding spacetime $\cM_E$ provided that $g\gg E$ for all $E$.

Hence, we conclude that the correspondence of space times $\cM_n$ or $\cM_E$ is actually realized for states close to the ground state for a strongly coupled scenario. Since $\cR_n,\cR_E\to 0$ as $\bar\a\to 0$,  $\cM_n$ or $\cM_E$  asymptotically behaves as a Rindler spacetime in that limit. This can been seen from 
\be
 \ud s^2= \ue^{\frac{4\xi}{L}}\left[-\frac{1}{1-\bar\a\ue^{\frac{2\xi}{L}}}\ud y^2 +\ud \xi^2\right]=\ue^{\frac{4\xi}{L}}\left[-\left(1+\bar\a\ue^{\frac{2\xi}{L}}+O(\bar\a^2)\right)\ud y^2 +\ud \xi^2\right].
 \label{expa}
\ee
The comparison between ground states is summarized in Table \ref{Table1}.
\begin{table}[h!]
\centering
 \begin{tabular}{|l|l|l|}
 \hline
Schr\"odinger CQM &Wheeler-DeWitt& Spacetime $\cM$\\
 \hline
&&\\
 $H\psi_{E=0}=0$ &$-\Delta \Psi=0$, $\Lambda=0$,  $g>0$ & Flat \\
 &&\\
 $R\varphi_{r_n=r_0}=0$ &$(-\Delta+1) \Psi=0$, $\Lambda=-\frac{1}{4}$, $g\gg 3$ &Negative curved\\
 &&\\
\hline
 \end{tabular}
 \caption{Spacetime realization of the $H$-vacuum and $R$-vacuum. The operator $\Delta$ is the Laplacian of the corresponding spacetime $\cM$.}
 \label{Table1}
\end{table}
The key ingredient for the comparison is the resulting Wheeler-DeWitt equation given in Eq. \eqref{ResultingWDW}. It corresponds to a Schr\"odinger equation with an effective potential of the form
\be
V_{\text{eff}}=\frac{g}{\bar{q}^2}-4\bar{\Lambda}\bar{q}^2,
\ee
and the eigenvalue is $\bar\epsilon=g\bar{\a}$. We see that only for negative $\bar\Lambda$ the potential is bounded from below. On the other hand, we are required to consider $\bar{\a}\to 0$ in order to fix the signature of the spacetime and this is achieved by considering the strong coupling limit. 

If we compare this results with the CQM model, we see that the issue of compactness of the operator translates into the mass/tension of the worldline propagating in $\cM$. The fact that the lowest energy eigenvalue of the compact operator is not zero corresponds to a non-vanishing parameter $\bar\a$ and consequently a deviation from Rindler spacetime. The existence of $\bar\a$ translates to curvature corrections to the flat case.

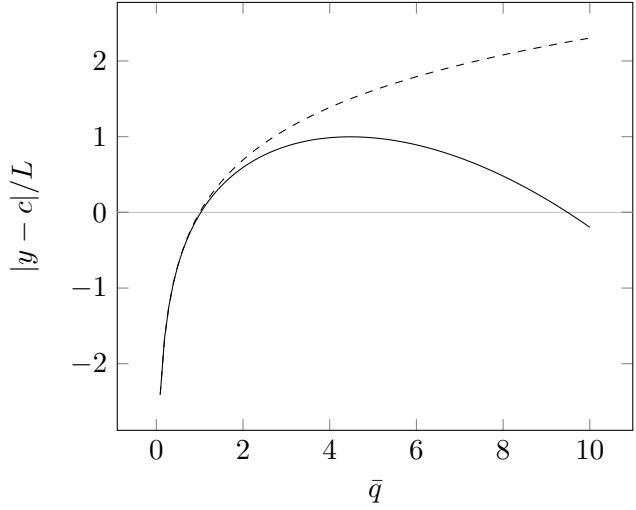
\begin{figure}[ht]
\centering
\begin{tikzpicture}
\begin{axis}[
    xlabel = $\bar{q}$,
   ylabel = {$|y-c|/L$},
   extra y ticks       = 0,
      extra y tick labels = ,
      extra y tick style  = { grid = major },
]

\addplot [
    domain=0.09:10, 
    samples=100, 
    dashed,
]
{ln(x)};

\addplot [
    domain=0.09:10, 
    samples=100, 
    color=black,
]
{ln(x)-(0.1/4)*x^2};
\end{axis}
\end{tikzpicture}
\caption{Plot of a null geodesics. The dashed curved corresponds to $\ln\bar{q}$ and the solid curve to $\bar{\a}=0.1$.}	
\label{NGPlot}
\end{figure}

For the ``physical'' spacetime we consider  small $\bar\a$ which corresponds to a negative curvature manifold. The null geodesics are 
\be
y=c\pm\left(L\ln\left(\frac{q}{L}\right)-\frac{\bar\a L}{4}\left(\frac{q}{L}\right)^2+O(\bar\a^2)\right). 
\ee
The logarithmic part corresponds to the 2-dimensional light cone. The $\bar\a$ ``deformation'' pulls the logarithmic curve down and produces a global maximum at $\frac{\sqrt{2}L}{\sqrt{\bar\a}}$, see Figure \ref{NGPlot}. 
The existence of this turning point is consistent with the  confinement property of $V\sim \frac{g}{q^2}+q^2$ in the CQM model. This feature is enjoyed for negative curve manifolds, e.g. we encounter similar behaviour of the geodesics in $AdS_2$.

Following the results of \cite{Claus:1998ts,BrittoPacumio:1999ax}, we are motivated to think that the $\bar\a$ ``deformations'' from Rindler arise from  corrections of the near-horizon geometry of a black hole and for this purpose let us consider a spherical symmetric non-rotating black hole solution in $d$ dimensions. The metric is given by 
 \be
 \ud s^2 = -f(r)\ud t^2+\frac{\ud r^2}{f(r)}+r^2\ud\Omega^2_{d-2}.
 \ee
 Let us assume that $f(r)=0$ has at least one real solution $r_h$. In the case of several real solutions, $r_h$ is taken to be the largest. Near the horizon we have $f(r)=f'(r_h)(r-r_h)+\frac{1}{2}f''(r_h)(r-r_h)^2+\ldots$. The sign of the correction to the flat metric is  determined by $f''(r_h)$. Notice that both $\ud t^2$ and $\ud r^2$ are modified.

As a first example consider Schwarzschild, i.e. $f(r)=1-\frac{2M}{r}$. We have $r_h=2M$, $f'(r_h)=\frac{1}{r_h}$ and $f''(r_h)=-\frac{2}{r_h^2}$. A second example is the charged black hole, i.e. $f(r)=1-\frac{2M}{r}+\frac{Q^2}{r^2}$. We obtain $r_h=M+\sqrt{M^2-Q^2}$ and
\be 
 f'(r_h)=\frac{2\sqrt{M^2-Q^2}}{(\sqrt{M^2-Q^2}+M)^2}, \quad f''(r_h)=\frac{6Q^2-4M(\sqrt{M^2-Q^2}+M)}{(\sqrt{M^2-Q^2}+M)^4}.
\ee
 Notice that for $\frac{2}{3}\sqrt{2}M<Q< M$ we have $f'(r_h),f''(r_h)>0$. From Einstein's equations of motion in $d=4$ we see that the Ricci scalar for the charged solution vanishes. This follows from the traceless of the electromagnetic energy-momentum tensor. Then,
\be
0=\cR_4=\cR_2+\frac{1}{r^2}\cR_{S^2}-\frac{2f(r)}{r^2}, 
\ee
or
\be
\cR_2=- \frac{2}{r^2}+\frac{2}{r^2}\left(1-\frac{2M}{r}+\frac{Q^2}{r^2}\right)=\frac{2Q^2}{r^4}-\frac{4M}{r^3},
\ee
where $\cR_2$ is the Ricci scalar of the 2d timelike submanifold. For  $r>\frac{Q^2}{2M}$ we obtain $\cR_2<0$ and $\cR_2\to 0$ as $r\to \infty$. Therefore, we find a near horizon correction of a charged black hole that is compatible with Eq. \eqref{expa} for $Q=\varepsilon M$ where 
$\frac{2}{3}\sqrt{2}<\varepsilon< 1$. Moreover, near the horizon the spacetime is negatively curved. Setting $r-r_h=\frac{1}{f'(r_h)}\ue^{f'(r_h)\xi}$ we obtain 
\be
\ud s_{\perp}^2=\ue^{f'(r_h)\xi}\left[-(1+\bar\a\ue^{f'(r_h)\xi}+\ldots)\ud t^2+(1-\bar\a\ue^{f'(r_h)\xi})\ud\xi^2\right], 
\label{CBH}
\ee
where 
\be
\bar\a=\frac{f''(r_h)}{2(f'(r_h))^2}=\frac{6\varepsilon^2-4(1+\sqrt{1-\varepsilon^2})}{8(1-\varepsilon^2)}. 
\ee
Notice that as $\varepsilon\to \frac{2}{3}\sqrt{2} $ we obtain $\bar\a\to 0$. After comparing Eq. \eqref{CBH} with Eq. \eqref{EqLineseg2} we see that the exponential function of the fine tune charged black hole scales differently from the ``deformed'' Rindler spacetime. Only for $\xi$ small we can claim that the $\bar{\a}$ ``deformation'' arises from curvature corrections of a charged black hole.

This result can be contrasted with the one obtained in \cite{Claus:1998ts}. The CQM model of \cite{D'Alfaro:693633} is obtained from an extreme Reissner-Nordstr\"om black hole with large mass. This follows basically from the fact that the near horizon geometry is AdS$_2\times S^2$ and the isometry group of AdS$_2$ is $SO(1,2)$\footnote{This also true for the 5-dimensional extreme Reissner-Nordstr\"om black hole dicussed in \cite{BrittoPacumio:1999ax}. In this case the near horizon geometry is AdS$_2\times S^3$. }. There are two major differences between the models: i) contrary to our case, the probe particle in  \cite{Claus:1998ts} is charged and can have angular momentum, ii) our black hole is not extreme ($\varepsilon\neq 1$).

Moving forward to another interpretation of the spacetime, we seek a solution of a two-dimensional dilaton gravity theory, see \cite{Grumiller:2002nm} for general models. Consider the following action
\be
 S=\frac{1}{2\kappa^2_{(2)}}\int\ud^2 x\sqrt{-G}\,\ue^{2\phi}\left[\cR-\frac{1}{2}\nabla_{\mu}\phi\nabla^{\mu}\phi\right]-M\int\ud\la\, \sqrt{-G_{\mu\nu}(Y(\la))\dot{Y}^{\mu}\dot{Y}^{\nu}}+ S_b,
\ee
where $S_b$ is the boundary action needed to have a well define first functional derivative of the action and $\kappa^2_{(2)}=8\pi G_N^{(2)}$. The equations of motion of the metric and dilaton are 
\bea
0&=&\frac{9}{2}\nabla_{\mu}\phi\nabla_{\nu}\phi +2\nabla_{\mu}\nabla_{\nu}\phi -G_{\mu\nu}\left(\frac{17}{4}\nabla_{\rho}\phi\nabla^{\rho}\phi+2\Box\phi \right)\nn
&&+\ue^{-2\phi}\kappa^2M\int	\ud\la\frac{\dot{Y}^{\rho}\dot{Y}^{\s}G_{\rho\mu}(Y(\la))G_{\s\nu}(Y(\la))}{\sqrt{-G_{\mu'\nu'}(Y(\la))\dot{Y}^{\mu'}\dot{Y}^{\nu'}}}\frac{\delta(x-Y(\la))}{\sqrt{-G(x)}},\label{eomDGBrane1}\\
0&=&\Box\phi+\nabla_{\rho}\phi\nabla^{\rho}\phi+2\cR, \label{eomDGBrane2}
\eea
respectively. Recall that in $d=2$ the Einstein tensor vanishes. In the weak gravity limit $\kappa^2_{(2)}\to 0$ the worldline decouples. The remaining equations can be combined to obtain
\be
 \frac{9}{2}\nabla_{\mu}\phi\nabla_{\nu}\phi +2\nabla_{\mu}\nabla_{\nu}\phi -G_{\mu\nu}\left(\frac{9}{4}\nabla_{\rho}\phi\nabla^{\rho}\phi-4\cR \right)=0.
\ee
Notice that for $\phi=\mathrm{constant}$ we obtain $\cR=0$. For a suitable solution we consider $\phi=\phi(q)$ together with  Eq.\eqref{EqLineseg2} and Eq. \eqref{riccia}. The equation  becomes
\be
\phi''+\frac{9}{8}(\phi')^2-\frac{1}{q}\phi'=\frac{4(4q_*^2-q^2)}{(q_*^2-q^2)^2},
\ee
where $q_*=\frac{R}{\sqrt{\bar\a}}$. For $q_*$ large and $q\ll q_*$ the source of the above equation vanishes and the solution is of the form
\be
\phi=c_1+\frac{8}{9}\ln\left( c_2+9\frac{q^2}{R^2}\right),\quad \phi'=\frac{16 q}{c_2R^2+9q^2}.
\ee
We choose $\phi(0)=0$, $\phi'(0)=0$ and  set $c_2=1$ ($c_1=0$).  As $q$ grows the solution becomes constant and we recover the flat case. We have found that the spacetime background can be obtained as a dynamical solution of a two-dimensional dilaton gravity model\footnote{In fact, the dilaton gravity model can be obtained from a dimensional reduction  of 
\be
S=\frac{1}{2\kappa_{(3)}^2}\int\ud^3 x\sqrt{-G_{(3)}}\,\cR_{(3)}+S_b, \quad \ud s^2_{(3)}= \ud s^2_{(2)}+\ue^{\sqrt{\frac{2}{3}}\phi(x)}\ud \chi,
\ee
where
\be
\frac{\Delta}{\kappa^2_{(3)}}=\frac{1}{\kappa^2_{(2)}}, \quad \Delta =\int\ud\chi.
\ee
Notice that for a non-compact extra-dimension, we obtain the two-dimensional weak limit.}.


\subsection{$|t\rangle$ basis and Dirac equation in 1+1}
We search for an interpretation of $|t\rangle$ basis from the geometrical point of view. From Eq. \eqref{t-rep-sl2al} and Eq. \eqref{Rt}, we see that the differential operators are linear in $\frac{\ud}{\ud t}$ rather then quadratic as for the previous cases. For this reason, we study the Dirac equation of a free massless fermion in a two-dimensional manifold. Following \cite{Koke:2016etw,Sabin:2016utj},  the metric is taken to be 
\be
 \ud s^2=\Omega^2(t,x)(\ud t^2-\ud x^2).
\ee
After a particular choice of gamma matrices, see \cite{Koke:2016etw,Sabin:2016utj}, the Dirac equation becomes
\be
 i\left(\pa_t+\frac{1}{2}\frac{\pa \ln \Omega}{\pa t} \right)\Psi=-i\s_x\left(\pa_x+\frac{1}{2}\frac{\pa \ln\Omega}{\pa x}\right)\Psi.
\ee
Let $\Omega$, $\Psi$ to be independent of $x$ and 
\be
\frac{1}{4}\left(t_*+\frac{t^2}{t_*}\right)\frac{\ud \ln \Omega}{\ud t}=r_0\frac{t}{t_*}+ir_n.
\ee
Then, from Eq. \eqref{Rt} we find $R\Psi_n=r_n\Psi_n$ and 
\be
\Omega_n(t)=\varpi_n\left(\frac{t_*-it}{t_*+it}\right)^{-2r_n}\left(1+\frac{t^2}{t_*^2}\right)^{2r_0}.
\label{omegasol}
\ee
After setting $\varpi_n=\frac{1}{C_n^2}$, where $C_n$ is given in Eq. \eqref{Cncoeff}, we find 
\be
 \langle t|n\rangle =\frac{1}{\Omega_n^{\frac{1}{2}}}, \quad \Psi_n(t)=\psi_0 \langle t|n\rangle,
 \label{spin}
\ee
where $\psi_0$ is a constant spinor. This result is equivalent to a local phase transformation of the spinor as discussed in \cite{Sabin:2016utj}. Hence, we find the correspondence 
\be
  \langle t|n\rangle\quad \longleftrightarrow\quad \ud s^2=\Omega_n^2(t)(\ud t^2-\ud x^2).
\ee
The spacetime corresponds to a complex manifold $\tilde{\cM}_n$ and this holds for every $n$. In this case the study of the geometry is more complicated since we are dealing with a complex spacetime. Nevertheless, the system provides another hint of the correspondence of a state in CQM and a manifold. 

From Eq. \eqref{sum1} and Eq. \eqref{2pfunc}, the two-point correlation function is given by the outer product
\be
\sum_{n=0}^{\infty}\Tr(\Psi_n(t_1)\Psi_n^{\dagger}(t_2))=\Tr(\psi_0\psi^{\dagger}_0)\frac{t_*^{2r_0}\G(2r_0)}{(2i(t_1-t_2))^{2r_0}}.
\ee
Notice that the left hand side is real and positive. Therefore we restrict $i^{2r_0}=e^{i\pi r_0}=1$. This implies that $r_0=2k$ where $k$ is a positive integer. Since we know that $r_0\geq \frac{3}{4}$, the restriction is always satisfied. The incompleteness of the $|t\rangle$-basis translates to
\be
 \int\limits_{-\infty}^{\infty}\ud t\, \Psi_m^{\dagger}(t)\Psi_n(t)=t_*C_{mn}(r_0) \psi^{\dagger}_0\psi_0,
 \label{incoferm}
\ee
where $C_{mn}(r_0)$ is given in Eq. \eqref{Cnm}.   The correlation functions defined in \cite{CHAMON2011503,Jackiw:2012ur} are
\be
 \sum_{n,m=0}^{\infty}\Tr(\Psi_m(t_1) \Psi^{\dagger}_n(t_N))\langle m| O_{\Delta}(t_2)\cdots O_{\Delta}(t_{N-1})|n\rangle=\Tr(\psi_0 \psi_0^{\dagger})\langle t_1|O_{\Delta}(t_2)\cdots O_{\Delta}(t_{N-1})|t_N\rangle,
 \ee
and the relation of the fermion with the  ``displacement'' operator is 
\be
\Psi^{\dagger}_n(t)=\psi_0^{\dagger}  \langle n|\cO(t)|0\rangle,
\ee
i.e., the matrix elements $ \langle n|\cO(t)|0\rangle$ are proportional to the adjoint spinor. Notice that the spinor can be written as 
\be
\Psi_n(t)=\psi_0 C_n \frac{\ue^{2ir_n\arctan\frac{t_*}{t }}}{(1+\frac{t^2}{t_*^2})^{r_0}}.
 \ee
Since the Dirac Lagrangian is invariant  under a global $U(1)$, the density   $\Psi^{\dagger}_n(t)\Psi_n(t)$ associated to this $U(1)$, peaks at $t=0$ and vanishes at $t\to\pm\infty$. The charge  for each $n$ can be computed from Eq. \eqref{incoferm} by setting $m=n$ in the expression.

On the other hand we have,
\be
 \langle n|\cO(t)|0\rangle=C_n\frac{\ue^{-2ir_n\arctan\frac{t_*}{t }}}{(1+\frac{t^2}{t_*^2})^{r_0}}.
\ee
The operator has a complex  $R$-vacuum expectation value. This is not a surprise since the operator is not primary (non-zero v.e.v.) and is not Hermitian (complex v.e.v.). The time dependence arises from the non-static nature of the spacetime. Recall that the matrix elements are related to $\Omega_n$ as shown in Eq. \eqref{spin}.

The spinor description of a gravitational bosonic system is not new. For example, in \cite{BenAchour:2020ewm} the cosmology of a free scalar can be reformulated in terms of a spinor. This cosmological spinor is constructed in phase space rather than configuration space as in our case. Another interesting example is the simulation of a   time dependent gravitational FRW background by a gauge potential in a Dirac equation as shown in \cite{1991Prama..36..519B}. We can also proceed in this line and define the charge $q$ and (complex) potential $\phi(t)$ such that $q\phi(t)=-\frac{i}{2}\frac{\pa\ln \Omega}{\pa t}$. From $\mathbf{E}=-\nabla\phi-\pa_t\mathbf{A}$, we see that there is no actual electric field.

Pursuing a different interpretation, we consider 
\be 
\tau_+=-2t_*\arctan \frac{t_*}{t }= \pi t_*-2 t_*\arctan \frac{t}{t_* },\quad t>0.
\ee
This is in the same line of  the $\tau$-time defined in \cite{Jackiw:2012ur}. The Dirac equation becomes
\be 
i\frac{\pa}{\pa\tau_+}\Psi^+_n=\frac{1}{t_*}\left[r_n+ir_0\cot\frac{\tau_+}{2t_*}\right]\Psi^+_n,
\ee
with
\be
\Psi^+_n(\tau_+)=\psi_0 C_n\left[\sin\frac{\tau_+ }{2t_*}\right]^{2r_0} \ue^{-ir_n\frac{\tau_+}{t_*}}, \quad 
\tau_+\in\left(-\pi t_*,0\right).
\ee
The spinor for $t<0$ can be constructed in a similar way. Consider the Floquet operator $\cH_F$ defined by\footnote{For  details on Floquet systems see e.g. \cite{2011PhRvB..84e4301A,2016JPhB...49a3001H,2020PhRvB.101m4302J,2020PhRvA.102d2212L} and references within.}
\be
\cH_F= H_n+V(\tau_+)-it_*\frac{\pa}{\pa\tau_+}, \quad H_n=r_n, \quad V(\tau_+)=ir_0\cot\frac{\tau_+}{2t_*},
\ee
where $H_n$ is the undriven Hamiltonian and $V$ is a non-hermitian time-periodic potential with period $T=4\pi t_*$. Thus, after extending the range of $\tau_+$, the Dirac equation can be rewritten as 
\be \cH_F \Psi^+_n(\tau_+)=0.
\ee 

Floquet's theory indicates that the eigenfunctions of  $\cH_F$ satisfy $\cH_F\Phi_{\a}(\tau_+)=\varepsilon_{\a}\Phi_{\a}(\tau_+)$ where  $\Phi_{\a}$ are called a Floquet modes   and $\varepsilon_{\a}$  the quasienergies\footnote{It is straightforward to see that quasienergies of two modes related by $\Phi_{\tilde\a}(\tau_+)=\Phi_{\a}(\tau_+)\ue^{i k\omega\tau_+}$ ($k\in\mathbb{N}$) can be expressed as $\varepsilon_{\tilde\a}=\varepsilon_{\a}+k\omega t_*$ where $\omega=\frac{2\pi}{T}$. This gauge transformation allow us to map the quasienergies into a first Floquet–Brillouin zone defined by $ - \frac{\omega t_*}{2}\leq \varepsilon\leq   \frac{\omega t_*}{2}$.}.  Floquet modes are periodic $\Phi_{\a}(\tau_++T)=\Phi_{\a}(\tau_+)$ and a general state can be written as $ \sum_{\a} c_{\a} \Phi_{\a}(\tau_+) \ue^{-i\varepsilon_{\a}\tau_+}$. Hence, the solutions of the Dirac equation corresponds to an infinite set of zero modes of the Floquet-Dirac eigenvalue problem\footnote{The zero modes are not orthogonal but they are normalizable. The norm is defined as 
\be
\|\Psi_n^{+}\|^2=\frac{1}{T}\int\limits_{0}^{T}\ud\tau_+\, \bar\Psi_n^{+}(\tau_+) \Psi_n^{+}(\tau_+) =\bar\psi_0\psi_0 C_n^2\frac{(1+(-1)^{4r_0})}{\sqrt{4\pi}}\frac{\Gamma[\frac{1}{2}+2r_0]}{\Gamma[1+2r_0]}.
\ee
}.
Notice that the periodic potential amplitude is $ir_0$. We have shown that $r_0\geq \frac{3}{4}$ and $r_0\neq 0$ imply that $|0\rangle$ is not a conformal vacuum. The later translate into the existence of the periodic potential in the Floquet-Dirac dual description.

Let us define the $\tau_+$-dependent operator  $U$ in the CQM model as
\be
U(\tau_+)=P(\tau_+)\ue^{i\bar{R}\frac{\tau_+}{t_*}}, 
\label{Ualmost}
\ee 
where $\bar{R}|n\rangle=r_n|n\rangle$. Then, $\langle n|U^{\dagger}(\tau_+)|0\rangle=\langle n|P^{\dagger}(\tau_+)|0\rangle \ue^{-ir_n\frac{\tau_+}{t_*}} $. If 
\be
 \langle n|P^{\dagger}(\tau_+)|0\rangle=C_n\left[\sin\frac{\tau_+ }{2t_*}\right]^{2r_0},
\ee
we find that $\cO(t(\tau_+))=U^{\dagger}(\tau_+)$ since 
 \be
  \langle n|\cO(t(\tau_+))|0\rangle=C_n\left[\sin\frac{\tau_+ }{2t_*}\right]^{2r_0} \ue^{-ir_n\frac{\tau_+}{t_*}}.
 \ee
Let  $|\la\rangle$ be a $\tau_+$-independent quantum state in the CQM model. Then, 
\be
\langle \la | U^{\dagger}(\tau_+)|0\rangle=\sum_n \langle \la|n\rangle C_n\left[\sin\frac{\tau_+ }{2t_*}\right]^{2r_0} \ue^{-ir_n\frac{\tau_+}{t_*}}.
\ee
The amplitude can be written in terms of the spinor as 
\be
 \langle \la | U^{\dagger}(\tau_+)|0\rangle\psi_0^{\dagger}=\sum_n \langle \la|n\rangle \Psi_n^{+\dagger}(\tau_+),
\ee
or 
\be
 \langle n | U^{\dagger}(\tau_+)|0\rangle\psi_0^{\dagger}=  \Psi_n^{+\dagger}(\tau_+).
\ee
Since $\cO$ is not unitary then $U$ is also not unitary and therefore it corresponds to an almost evolution operator of a periodic system. The form of $U$ suggests the description of an almost Floquet-CQM system. The non-unitarity of $U$ can be also seen from the expression
\be
\langle 0|P(\tau_+)P^{\dagger}(\tau_+)|0\rangle=\left[\sin\frac{\tau_+ }{2t_*}\right]^{4r_0}\sum_n C_n^2.
\ee
For a true Floquet system $P$ is unitary. The above expression indicates that this can not be true. This maybe expected if we translate the non-unitary of $P$ in the CQM model  to the non-hermitian periodic potential in the gravitational Floquet-Dirac system.

Within the Floquet-Dirac system interpretation, the questions raised in the introduction can be answered as follows:
\begin{itemize}
\item The gravitational dual   is based on the  Dirac equation of a massless spinor in a two-dimensional curved background. The classical action associated to this equation is Weyl invariant and therefore the conformal structure is expected. The $R$-vacuum and $\cO$ appear in the spinor solutions as a matrix element $\langle 0|\cO^{\dagger}(t)|n\rangle$. Moreover, they appear in the $N$-point correlation function via the outer product $\Tr(\Psi_m(t_1) \Psi^{\dagger}_n(t_N))$.
\item  The spinor solutions can be interpreted as  zero modes of a Floquet-Dirac system. A Floquet-CQM system fails to be realized since the would-be evolution operator is not unitary. The adjoint of this operator corresponds to the ``displacement'' operator $\cO$. 
\end{itemize}

\section{The general dual proposal and conclusions}
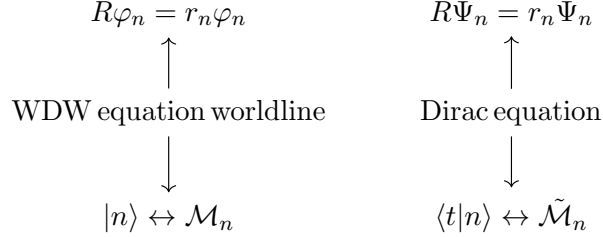
\begin{figure}[ht]
\[
\begin{tikzcd}
R\varphi_n=r_n\varphi_n  & R\Psi_n=r_n\Psi_n\\
\mathrm{WDW\,equation\,worldline}   \arrow[swap]{u} \arrow[swap]{d}  & \mathrm{Dirac\,equation}   \arrow[swap]{u} \arrow[swap]{d} \\
 \vert n\rangle \leftrightarrow\mathcal{M}_n & \langle t\vert n\rangle\leftrightarrow\tilde{\mathcal{M}}_n
\end{tikzcd}
\]
\caption{Correspondence of a state in CQM and a 2d manifold.}
\label{geoex}
\end{figure}

The geometrical exploration of the last section is summarized in Figure \ref{geoex}. 
These two descriptions share the correspondence of state-geometry with different dynamics, worldline versus fermion. One may attempt to unify the description by considering fields only. For the $|q\rangle$ basis representation of the $R$ operator, this would imply that the equation of motion of a scalar field in a static background should correspond to the eigenvalue problem of $R$. If we consider a system of  a free massless scalar and free massless Dirac spinor in a fixed background, we will run into the problem that the equation for the scalar is reduced to the wave equation in flat spacetime. Moreover, since gravity alone is topological in two dimensions we need to consider a dilaton gravity  again for a dynamical model. Therefore, we propose the following action
\be
 S_{d=2\,\mathrm{DG}+\mathrm{F}}=\frac{1}{2\kappa^2_{(2)}}\int\ud^2 x\sqrt{-G}\,\ue^{2\phi}\left[\cR-\frac{1}{2}\nabla_{\mu}\phi\nabla^{\mu}\phi-\ue^{-2\phi}V(\phi)+i\ue^{-2\phi}\bar\Psi\slashed\nabla\Psi\right]+ S_b,
\ee
Let us consider the ansatz 
\be
\ud s^2=\Omega^2(t)(-\ud t^2+\ud x^2), \quad \phi=\phi(x),\quad  V(\phi)=0,\quad \Psi=\Psi(t).
\ee
The action becomes 
\be
 S_{d=1\,\mathrm{F}}=\frac{1}{2\kappa^2_{(2)}}\int\ud t\left[L_0 U(\Omega.\dot{\Omega},\ddot{\Omega})-\frac{iL_1}{\Omega}\Psi^{\dagger}\left(\frac{\ud }{\ud t}+\frac{1}{2}\frac{\ud}{\ud t}\ln\Omega\right)\Psi\right]+ S_b,
\ee
where $U$ is a function and 
\be
L_0=\int\limits_{a}^{b} \ud x\, \ue^{2\phi(x)},\quad L_1=\int\limits_{a}^{b} \ud x= b-a.
\ee
Since $\phi(x)$ is not dynamic we are free to choose it such that $L_0\to 0$ and therefore the resulting dimensionally reduced action describes a time dependent fermion with equation of motion $i\left(\frac{\ud }{\ud t}+\frac{1}{2}\frac{\ud}{\ud t}\ln\Omega\right)\Psi=0$. Let $x$ correspond to an angular variable, then  $x=L\theta$ where $L$ is an intrinsic scale. We assume that $b=\frac{\pi}{2}$ and $a=-\frac{\pi}{2}$. Thus, $\ue^{2\phi(L\theta)}$ must be odd in the domain $[-\frac{\pi}{2},\frac{\pi}{2}]$. Hence, the spacetime is conformally related to the Lorentzian cylinder of radius $L$. The conformal factor is given by Eq. \eqref{omegasol}.
 
 Consider another ansatz
 \be
\ud s^2=A^2(x)(-\ud t^2+\ud x^2), \quad \ue^{\phi(x)}= \sqrt{\frac{2\kappa_{(2)}^2}{\Delta t}}\varphi(x) , \quad V(\phi)=\frac{m^2}{2}\ue^{2\phi},\quad \Psi=0,
\ee
where $\Delta t=\int\ud t$. The action becomes
\be
S_{d=1\,\mathrm{D}} =\int\ud x\left[-\frac{1}{2}(\varphi')^2-\frac{A^2m^2}{2}\varphi^2\right]+S_b.
\ee
 The equation of motion is $\varphi''-A^2m^2\varphi=0$. Let us introduce the dimensionless coordinate $\bar{x}=mx$ and consider
 \be
 A^2= \frac{g}{\bar{x}^2}+\lambda\bar{x}^2-4r_n,
 \label{cfA}
 \ee
 where $\lambda\geq 0$ is a dimensionless parameter. The resulting equation becomes
 \be
 \frac{1}{4}\left[-\frac{\ud^2}{\ud \bar{x}^2}+\frac{g}{\bar{x}^2}+\lambda\bar{x}^2\right]\varphi_n=r_n\varphi_n.  
 \ee
 For $\la=1$ we recover the $R$ operator in the position basis. The spacetime is conformally related to the Lorentzian strip of width $\Delta t$.
 
\begin{figure}[ht]
\centering
\begin{tikzpicture}[node distance=2cm]
\node (2d) {$S_{d=2\,\mathrm{DG}+\mathrm{F}}$};
\node (CQM) [below of=2d, xshift=0cm, yshift=-2.6cm] {CQM};
\node (FD2) [below of=2d, xshift=0cm, yshift=-1cm] {$\cH_F\Psi_n=0$};
\node (2dF) [below of=2d, xshift=3.3cm, yshift=0.5cm] {$S_{d=1\,\mathrm{F}}$};
\node (2dD) [below of=2d, xshift=-3.3cm, yshift=0.5cm] {$S_{d=1\,\mathrm{D}}$};
\node (Rt) [below of=2dF, yshift=0.5cm] {$ R\Psi_n=r_n\Psi_n$};
\node (Rq) [below of=2dD, yshift=0.5cm] {$ R\varphi_n=r_n\varphi_n$};
\node (n) [below of=Rq,, yshift=0.5cm,xshift=-1cm]{$\langle q|n\rangle\Leftrightarrow A^2(\mathrm{Lorentzian\, strip)}$};
\node (tn) [below of=Rt,, yshift=0.5cm,xshift=1cm]{$\langle t|n\rangle\Leftrightarrow \Omega^2(\mathrm{Lorentzian \,cylinder})$};
\draw [arrow] (2d) -- (2dF);
\draw [arrow] (2d) --  (2dD);
\draw [arrow] (2dF) -- (Rt);
\draw [arrow] (2dD) -- (Rq);
\draw [arrow] (Rq) -- (n);
\draw [arrow] (Rt) -- (tn);
\draw [arrow,<-] (Rq) -- (CQM);
\draw [arrow] (CQM) -- (Rt);
\draw [arrow,<->] (FD2) -- (Rt);
\end{tikzpicture}
\caption{General dual model.}
\label{dualmodel}
\end{figure}
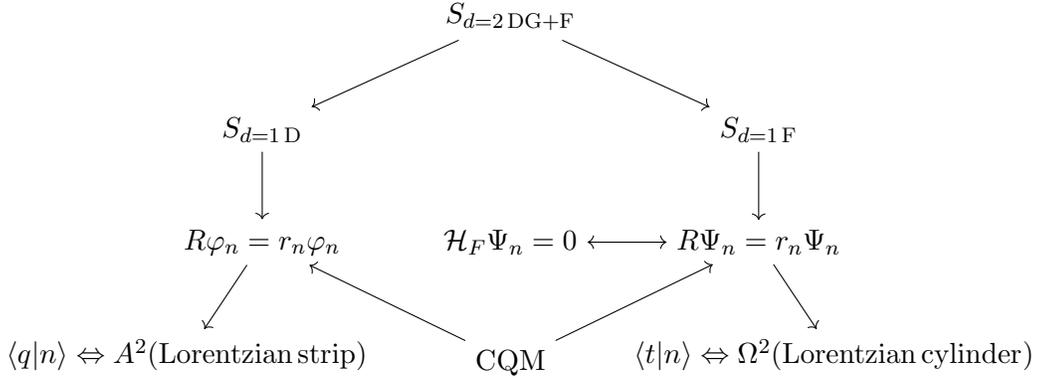

We see that  the $R$ operator in each corresponding basis is recovered from the conformal factor. The general dual model is shown in Figure \ref{dualmodel}.
We conclude that this is a universal correspondence between CQM states and geometries, c.f. with Figure \ref{geoex}. More importantly than the  CQM/geometry correspondence is the realization of the fermion as a Floquet-Dirac system. The key points of this interpretation are:
\renewcommand{\labelenumi}{\roman{enumi})}
 \begin{enumerate}
\item The non-vanishing lowest weight $r_0$ indicates that  $R$-vacuum  is not conformally invariant and also indicates the presence of the non-Hermitian periodic potential $V$ in the Floquet-Dirac system.
\item The inequality $r_0\geq\frac{3}{4}$ is satisfied for a strong coupling CQM. 
\item The adjoint of the ``displacement'' operator, i.e. $\cO^{\dagger}$, corresponds to a non-unitary $\tau_+$-time evolution operator in the CQM. 
\item We can interpret the non-unitarity as the consequence  non-Hermitiancy of $V$. This prevents the realization of a Floquet-CQM system. 
\item The conformal structure follows from the Weyl invariance of the Dirac action in two-dimensions.
\end{enumerate}
The relation of the Floquet-Dirac system with a scalar in  AdS$_2$ remains to be studied in order to have a complete understanding of AdS$_2$/CQM. 

Finally, as a product of the dual proposal, we notice that the Floquet-Dirac equation $\cH_F\Psi=0$ may be interpreted as the dual of $\Omega\psi=0$ with $\Omega=i\pa_t+\frac{1}{2}\pa_q^2-\frac{g}{2}\frac{1}{q^2}-\frac{1}{2}q^2$. As discussed in \cite{Valenzuela:2009gu,Toppan:2014kqa,Toppan:2017zxh} and references within, for a constant, linear and quadratic potential the corresponding equation $\Omega\psi=0$ is invariant under the Schr\"odinger algebra $\mathfrak{sch}(1)$ in the presence of a central charge. Therefore, for the $1/q^2+q^2$ potential, $\Omega\psi=0$ is invariant under a subalgebra of $\mathfrak{sch}(1)$. Moreover, as conjectured in \cite{Valenzuela:2009gu} and further supported for an interacting non-relativistic particle in \cite{Toppan:2014kqa}, there is a connection between the infinite symmetries of the free Schr\"odinger equation and higher spin symmetries. The infinite symmetries for the Floquet-Dirac equation should correspond to the gauge symmetry of the quasienergies. This will be addressed in future work.

\acknowledgments
I would like to thank Claudette Dawood and Alberto Faraggi for useful discussions. Also to Francesco Toppan and Mikhail Plyushchay for pointing out useful references. This work was funded by the National Agency for Research and Development (ANID), Concurso FONDECYT de Postdoctorado 2020 \# 3200721.

\appendix
\section{Self-adjoint test of $R$}
\label{appA}
Following \cite{doi:10.1119/1.1624111,doi:10.1119/1.2165248}, we need to solve 
\be
 \frac{1}{4}\left[-\frac{\ud^2}{\ud q^2}+\frac{g}{q^2}+q^2\right]\varphi_{\pm}=\pm i\eta \varphi_{\pm},
\ee
where $\eta$ is real and positive. The solutions that vanish at the origin are 
\be
\varphi_{\pm}(q)=C_{\pm}\ue^{\mp \frac{q^2}{2}} q^{\frac{1}{2}+\frac{1}{2}\sqrt{1+4g}}L^{\frac{1}{2}\sqrt{1+4g}}_{i\eta-\frac{1}{4}\sqrt{1+4g}-\frac{1}{2}}(\pm q^2),
\ee
where $L$ is associated Laguerre function. Using
\begin{gather}
 L^{\frac{1}{2}\sqrt{1+4g}}_{i\eta-\frac{1}{4}\sqrt{1+4g}-\frac{1}{2}}(\pm q^2)=f(g,\eta)\nn
\times \, _1F_1(-i\eta+\frac{1}{4}\sqrt{1+4g}+\frac{1}{2},\frac{1}{2}\sqrt{1+4g}+1,\pm q^2),\nn
 f(g,\eta)=\frac{\G(\frac{1}{2}\sqrt{1+4g}+1)}{\G(\frac{1}{2}\sqrt{1+4g})\G(i\eta-\frac{1}{4}\sqrt{1+4g}+\frac{1}{2})},
\end{gather}
the behaviour as $q\to\infty$ result
\be
 L^{\frac{1}{2}\sqrt{1+4g}}_{i\eta-\frac{1}{4}\sqrt{1+4g}-\frac{1}{2}}(\pm q^2)\sim \ue^{\pm q^2}q^{-2i\eta-2r_0}.
\ee
Then, 
\be 
\varphi_{\pm}(q)\sim \frac{\ue^{-2i\eta\ln q}}{\sqrt{q}},
\ee
 in this limit. We conclude that $\varphi_{\pm}$ are not normalizable and therefore $R$ is self-adjoint.

\bibliographystyle{JHEP}
\bibliography{SCQM} 
\end{document}